\documentclass[10pt,conference]{IEEEtran}
\IEEEoverridecommandlockouts
\def\BibTeX{{\rm B\kern-.05em{\sc i\kern-.025em b}\kern-.08em
    T\kern-.1667em\lower.7ex\hbox{E}\kern-.125emX}}

\usepackage{graphicx}
\usepackage{textcomp}
\usepackage{xcolor}
\usepackage[most]{tcolorbox}
\usepackage{etoolbox}
\usepackage{colortbl}
\usepackage{tikz}
\usepackage{url}
\usepackage{environ}
\usepackage{balance}
\usepackage{hyperref}
\usepackage{multirow}
\usepackage{pifont}
\usepackage{adjustbox}
\usepackage{booktabs}
\usepackage{utfsym}
\usepackage{xcolor}

\newcommand{\greencheck}{\textcolor[RGB]{40,184,18}{\usym{1F5F8}}}%

\newcommand{\redcross}{\textcolor{red}{\usym{2717}}}

\title{The Dark Art of Financial Disguise in Web3: Money Laundering Schemes and Countermeasures}

\author{
\IEEEauthorblockN{Hesam Sarkhosh}
\IEEEauthorblockA{
University of Waterloo \\
hsarkhos@uwaterloo.ca}
\and
\IEEEauthorblockN{Uzma Maroof}
\IEEEauthorblockA{
University of Waterloo \\
uzma.maroof@uwaterloo.ca}
\and
\IEEEauthorblockN{Diogo Barradas}
\IEEEauthorblockA{
University of Waterloo \\
diogo.barradas@uwaterloo.ca}
}


\newcommand{\todoc}[1]{\textcolor{red}{\textbf{TODO} - \small\textit{#1}}}

\newcommand{\mypara}[1]{\vspace{2pt}\noindent{\bf{#1}}}
\newcommand{\myparait}[1]{\vspace{2pt}\noindent{\textit{#1}}}

\renewcommand{\todoc}[1]{}

\newif\ifhighlight
\highlightfalse
\ifhighlight
  
\else
  
\fi

\begin{document}

\maketitle

\begin{abstract}
The rise of \textit{Web3} and \textit{Decentralized Finance (DeFi)} has enabled borderless access to financial services empowered by smart contracts and blockchain technology. However, the ecosystem's trustless, permissionless, and borderless nature presents substantial regulatory challenges. The absence of centralized oversight and the technical complexity create fertile ground for financial crimes. Among these, \textit{money laundering} is particularly concerning, as in the event of successful scams, code exploits, and market manipulations, it facilitates covert movement of illicit gains. Beyond this, there is a growing concern that cryptocurrencies can be leveraged to launder proceeds from drug trafficking, or to transfer funds linked to terrorism financing.

This survey aims to outline a taxonomy of high-level strategies and underlying mechanisms exploited to facilitate money laundering in Web3. We examine how criminals leverage the pseudonymous nature of Web3, alongside weak regulatory frameworks, to obscure illicit financial activities. Our study seeks to bridge existing knowledge gaps on laundering schemes, identify open challenges in the detection and prevention of such activities, and propose future research directions to foster a more transparent Web3 financial ecosystem---offering valuable insights for researchers, policymakers, and industry practitioners.

\end{abstract}

% Color and styling
\colorlet{verylightgrey}{gray!15}

\section{Introduction} 
\label{sec:introduction}

Blockchains emerged as a transformative technology aimed at decentralization, transparency, and trustless transactions. This vision materialized in 2009 with the launch of \textit{Bitcoin}~\cite{nakamoto2008bitcoin}, the first decentralized digital currency enabling peer-to-peer (P2P) transfers without intermediaries. Bitcoin marked a paradigm shift, replacing institutional trust with cryptographic proof~\cite{bellaj2022sok}. The second major milestone was Ethereum~\cite{buterin2014next}, which introduced smart contracts, expanding blockchains into programmable platforms for decentralized applications (\textit{dApps})~\cite{bellaj2022sok}. As adoption grew, so did the broader idea of decentralized systems, culminating in the concept of \textit{Web3}, coined %by Ethereum co-founder Gavin Wood 
in 2014, to describe a decentralized alternative to the traditional internet~\cite{edelman_father_nodate}. Web3's emergence catalyzed the rise of \textit{decentralized finance (DeFi)}---a blockchain-based, P2P financial system offering alternatives to traditional mechanisms such as lending and investment~\cite{wernerSoKDecentralizedFinance2023}. As of July 2025, the global cryptocurrency market cap reached \$3.8 trillion, with DeFi accounting for \$135 billion in total value locked (TVL), demonstrating DeFi's expanding global impact~\cite{coingecko2025,defillama2025}.

Critics have long warned that the unregulated nature of Web3 facilitates illicit activity~\cite{wernerSoKDecentralizedFinance2023}, including fraud, scams, wash trading, and code exploits~\cite{wu2023financial}. The presence of high-value assets continues to attract cybercriminals; in 2024 alone, \$2.2 billion was stolen from Web3 platforms~\cite{chainalysisStolen2025}, followed by the recent \$1.5 billion \textit{Bybit} hack in February 2025---considered the largest crypto theft to date~\cite{reutersBybitHack2025}. A key enabler of such crimes is the ability to obscure the origin of illicit assets, allowing launderers to integrate them into the legitimate economy---a process known as \textit{money laundering}~\cite{treasury_money_2024}. Frequently associated with drug trafficking and terrorism financing, money laundering is increasingly leveraging Web3~\cite{wu2023financial,kolachala2021sok,comolli_surfing_2021}; between 2019 and 2023, an estimated \$93 billion was laundered via cryptocurrencies~\cite{chainalysis_report_2024}.

Despite growing concern, crypto money laundering schemes remain highly diverse and insufficiently systematized. While works such as Sneck~\cite{sneck2024cryptocurrency}, Comolli and Korver~\cite{comolli_surfing_2021}, and Almeida et al.~\cite{almeida2023review} provide partial overviews, they often lack depth, breadth, or up-to-date coverage. Foundational studies like Möser et al.~\cite{moser_inquiry_2013}, though influential, are now outdated and omit recent developments. Even more recent contributions, such as Lin et al.~\cite{lin2023towards}, fall short of offering a comprehensive taxonomy of laundering schemes. A parallel body of work examines only specific laundering vectors---such as \textit{mixing services}~\cite{shojaeinasab_decoding_2024,arbabi2023mixing,ruffing2017p2p,crawford2020knowing}, \textit{initial coin offerings}~(ICOs)~\cite{barone2019cryptocurrency}, \textit{cross-chain bridges}~\cite{hu2024piecing}, and \textit{non-fungible tokens}~(NFTs)~\cite{wu2023financial,mooij2024regulating}. While informative, these studies remain fragmented and lack synthesis across the broader laundering landscape. 

For addressing the above gap, this work delivers a survey that consolidates known cryptocurrency-based money laundering approaches identified in academic and regulatory sources. We categorize these approaches by their overarching strategies and the tools or intermediaries employed, and map them to the respective stages of the laundering process. Recognizing the role of data in anti-money laundering efforts, we also examine the types and sources of data relevant to each approach. We provide a comparison with previous surveys in Appendix~\ref{appendix:literature-comparison}.

We begin by outlining the foundations of Web3 and DeFi~(\S\ref{sec:background}) and describing our methodology for data collection and analysis~(\S\ref{sec:methodology}). We then examine how traditional laundering stages adapt to Web3 and the challenges of deanonymization in this emerging ecosystem~(\S\ref{sec:ml-primer}). Next, we present our taxonomy of laundering schemes~(\S\ref{sec:ml-schemes}), including general obfuscation strategies~(\S\ref{subsec:ml-strategies}) and intermediary mechanisms exploited throughout the laundering process to implement these strategies~(\S\ref{subsec:ml-intermediaries}). We conclude with a survey of countermeasures from academia, regulators, and industry~(\S\ref{sec:ml-defenses}), and outline open challenges and future research directions~(\S\ref{sec:future-directions}).

\section{Foundations of Web3 and DeFi}
\label{sec:background}
This section outlines foundational concepts essential for understanding Web3 and DeFi. Key topics include the underlying blockchain infrastructures, classification of digital assets, and the role of crypto wallets in asset management. Understanding how value is created, stored, and moved in this ecosystem is critical for assessing its susceptibility to money laundering.

\mypara{Blockchains and consensus.}
A blockchain is a decentralized, cryptographically linked ledger maintained across a P2P network~\cite{xu2023survey,kwon2019eye}. To ensure consistency and \textit{Byzantine Fault Tolerance}~(BFT)---resilience to malicious parties---blockchains rely on consensus protocols~\cite{xu2023survey}. Bitcoin introduced the \textit{Nakamoto consensus}~\cite{nakamoto2008bitcoin}, using \textit{proof of work (PoW)} to achieve consensus by requiring nodes to solve computational puzzles; the longest valid chain is treated as authoritative~\cite{xu2023survey,kwon2019eye}. In contrast, Proof of Stake (PoS) protocols select block proposers based on the amount of possessed and locked assets (stake), reducing energy use by eliminating intensive computations. Both models incentivize participation through transaction fees and block rewards, and aim to maintain integrity and immutability.

\mypara{Smart contracts.}
Conceptualized by Szabo in the 1990s~\cite{szabo1997formalizing}, smart contracts are self-executing programs that enforce predefined rules without intermediaries~\cite{wu2023financial}. Unlike traditional contracts, enforcement is automatic and encoded within the contract logic~\cite{bonneau_sok_2015,wernerSoKDecentralizedFinance2023}. Ethereum~\cite{buterin2014next} extended blockchains into programmable state machines, enabling Turing-complete logic via smart contracts~\cite{li2022sok}. These contracts execute atomically---either fully completing or reverting without side effects~\cite{wernerSoKDecentralizedFinance2023}. Smart contracts are the foundation of Web3, powering decentralized applications (\textit{dApps}) and asset exchange~\cite{adamyk2025risk, wu2023financial,wernerSoKDecentralizedFinance2023}.

\mypara{Crypto assets.}
We adopt the term \emph{crypto assets} as a unified descriptor for blockchain-based, exchangeable digital assets~\cite{finraCryptoAssets2024,xu2023sok}, where ``crypto'' denotes cryptocurrency. While alternative terms such as ``digital assets'' and ``crypto-coins (CCs)'' are also prevalent~\cite{wernerSoKDecentralizedFinance2023,campbell2018bitcoin}, we use \emph{crypto assets} throughout for consistency. This category includes:

\myparait{Coins and gas tokens.}
Blockchains such as Bitcoin~\cite{nakamoto2008bitcoin}, Monero~\cite{monero}, and Ripple~\cite{xrpledger} support a single native cryptocurrency---BTC, XMR, and XRP---commonly referred to as ``coins,'' used as digital currency and for paying transaction or mining fees~\cite{mohit2022coin,xu2023survey}. In contrast, smart contract platforms like Ethereum~\cite{buterin2014next}, Solana~\cite{solana}, and Binance Smart Chain~\cite{bsc} use native ``gas tokens''---ETH, SOL, and BNB--- to cover transaction costs and smart contract execution~\cite{xu2024exploring}.

\myparait{Utility and community tokens.}
Blockchains with smart contract support enable the creation of secondary tokens via standards such as ERC-20~\cite{erc20Standard} in Ethereum and BEP-20~\cite{bep20Standard} in Binance Smart Chain. These tokens may serve functional roles (e.g., governance, staking) or may lack explicit utility~\cite{21shares2023classification, semanticScholarTokens2022}. 

\myparait{Wrapped tokens.}
These are representations of blockchain-native assets that enable interoperability across chains. Since assets cannot move directly between blockchains due to isolated state machines, bridges facilitate wrapping by locking the original asset on the source chain and minting an equivalent token on the destination chain~\cite{ou2022overview,lee2023sok}. For instance, locking BTC can yield wBTC for use in Ethereum-based protocols~\cite{lee2023sok}. Bridge protocols generally consist of three parts: a \textit{custodian} that locks the asset, a \textit{communicator} that transmits confirmations, and a \textit{debt issuer} that mints or burns the wrapped token~\cite{lee2023sok}.

\myparait{Stablecoins.}
Stablecoins are digital assets pegged to external references (typically fiat, e.g., USD)~\cite{moin2020sok,mooij2024regulating}. Unlike volatile cryptocurrencies like Bitcoin, they aim to serve as stable stores of value~\cite{moin2020sok}. Stablecoins are classified by~\cite{moin2020sok}: peg type (fiat, commodities~\cite{coinbaseRWA2024}), collateral type (fiat, crypto, commodities, algorithmic), collateral amount (full, over-collateralized, partial, or none), and stabilization mechanisms. Their stability depends on collateral volatility, reserves, macroeconomic conditions, and user incentives~\cite{kwon2023drives,hafner2023four}.

\myparait{Non-fungible tokens (NFTs).}  
Web3 tokens are broadly classified as \textit{fungible} or \textit{non-fungible}~\cite{wu2023financial}. Fungible tokens are interchangeable and divisible, with each unit holding equal value. In contrast, NFTs are unique, indivisible assets that function as digital certificates of ownership~\cite{mooij2024regulating}. Each NFT is identified by an ID and linked to off-chain metadata (e.g., media files) via a URL~\cite{wu2023financial_chapter1}. NFTs serve as foundational components of Web3 metaverse ecosystems by facilitating verifiable, transferable ownership of virtual assets~\cite{mooij2024regulating,white2022characterizing,wu2023financial}. Unlike fungible tokens with market-driven pricing, NFT values are subjective~\cite{upadhyay2025dark}, often exchanged through auctions on platforms like OpenSea~\cite{white2022characterizing}. 
% Their subjective valuation make NFTs particularly vulnerable to misuse (see~\S\ref{subsec:ml-strategies}).

\mypara{Cryptocurrency wallets.}
Cryptocurrency wallets are essential tools for storing, managing, and transferring digital assets within virtual economies~\cite{mooij2024regulating}. A wallet enables users to initiate transactions by signing them with a private key, and broadcasts it to the blockchain for validation~\cite{houy2023security}. Wallets vary by design and provider~\cite{mooij2024regulating}, and are typically classified along two dimensions~\cite{houy2023security,mooij2024regulating}: (a) network connectivity, where hot wallets (e.g., web, mobile) are internet-connected, and cold wallets (e.g., hardware, paper) are offline to reduce exposure to cyber threats; and (b) key management, where custodial wallets entrust private keys to third parties, while non-custodial wallets give users full control and responsibility. Cryptocurrency wallets are typically initialized using a 12- or 24-word mnemonic phrase drawn from a standardized 2048-word list~\cite{mnemonics-bip39,houy2023security}. These phrases encode entropy and act as human-readable seeds for private key generation, enabling secure recovery~\cite{mnemonics-bip39,houy2023security}. Most wallets implement the hierarchical deterministic (HD) model defined in BIP-32~\cite{hdwallets-bip23}, allowing a single master seed to produce a tree of key pairs~\cite{courtois2017stealth}. Protocol-specific derivation paths~\cite{bip44} yield distinct addresses, and the one-way nature of key derivation makes seed reconstruction or cross-chain address inference cryptographically infeasible. The ease of wallet generation facilitates abuse, allowing malicious parties to create large numbers of wallets~\cite{mooij2024regulating}. Some networks, such as Ripple~\cite{xrpledger}, disincentivize mass wallet generation by imposing a reserve balance requirement.

\section{Survey Methodology}
\label{sec:methodology}

In our survey, we primarily used Google Scholar to gather research from venues in cybersecurity, blockchain, and financial crime. Given the rapid evolution of DeFi, academic literature alone is insufficient for capturing emerging laundering tactics. To supplement scholarly work, we included regulatory reports, preprints, books, and theses, spanning the intersection of topics between finance and computer security. We also drew on curated SoK resources~\cite{hookleeSoKLibrary,OaklandSoK} and contextualized real-world incidents via reputable news sources and blogs.

We queried Google Scholar using keyword sets such as \textit{``Cryptocurrency money laundering,'' ``AML in Web3,'' ``crypto laundering,'' ``DeFi money laundering,'' ``mixing services,''} and related terms. These searches yielded over 50 published academic papers and preprints. Key articles were selected based on relevance, citation count, and venue, and forward citation analysis was used to identify additional sources. Given our work's interdisciplinary scope, we expanded our search to finance literature via our university's library, guided by a subject librarian. This resulted in the finding of additional studies, expanding our reference pool to over 150 sources. We shortlisted articles based on titles, followed by screening abstracts and introductions. Final selections were made after reviewing core technical sections. Many of these works cited non-peer-reviewed sources, including publications from industry, financial watchdogs (e.g., FinCEN, Interpol), and annual reports from forensic firms such as Chainalysis~\cite{chainalysis}, Elliptic~\cite{elliptic}, Mandiant~\cite{mandiantWebsite}, and Merkle Science~\cite{merklescience}. 

\begin{figure*}[t]
\centering
\includegraphics[width=0.99\textwidth]{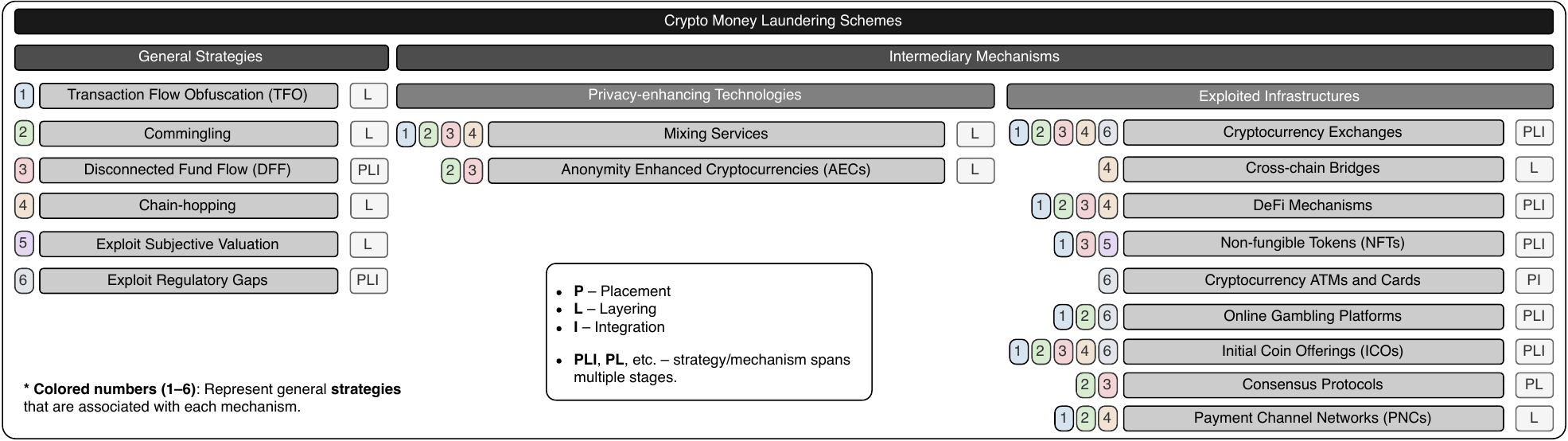}
\vspace{-0.2cm}
\caption[Taxonomy of crypto money laundering schemes]{
    Taxonomy of crypto money laundering schemes in light of the placement, layering, and integration (PLI) stages.}
    \vspace{-0.3cm}
\label{fig:taxonomy}
\end{figure*}

To construct our taxonomy, we began by listing all laundering approaches identified in the literature, supplementing each with domain-specific sources to enhance coverage and accuracy. Then, we derived a top-down model that organizes techniques into six overarching strategies, each supported by specific intermediaries and enablers grouped into two categories and eleven distinct elements~(see \autoref{fig:taxonomy}). To ensure completeness, we extended our investigation beyond academic literature, identifying emerging techniques not previously documented---see Appendix~\ref{appendix:literature-comparison}. We also synthesized countermeasures from academic, regulatory, and industry sources, classifying prevention and detection strategies by approach and scope to complement our analysis and provide an overview of defenses aimed at tackling crypto money laundering.

\section{Primer on Crypto Money Laundering}
\label{sec:ml-primer}

Money laundering is the process of concealing the illicit origins of criminal proceeds to make them appear legitimate~\cite{wu2023financial,comolli_surfing_2021}. By obscuring financial trails, money laundering enables offenders to integrate unlawful gains into the legal economy while concealing the underlying crimes and perpetrators~\cite{treasury_money_2024,FBIMoneyLaundering}. This often involves transforming assets, transferring them across jurisdictions, or layering transactions~\cite{comolli_surfing_2021}. A big concern with Web3 is its exploitation for money laundering~\cite{sneck2024cryptocurrency}. The pseudonymity of cryptocurrency transactions hampers regulatory and law enforcement efforts by obscuring the origin/destination of funds. Criminals often favor decentralized services over centralized ones with AML controls~\cite{treasury_money_2024}. Crypto money laundering is also used to integrate proceeds from Web3-native crimes such as scams and hacks~\cite{comolli_surfing_2021,wu2023financial}.

\mypara{Money laundering stages.}
The money laundering process is traditionally divided into three primary stages: placement, layering, and integration (PLI)~\cite{comolli_surfing_2021,sneck2024cryptocurrency,kolachala2021sok,mooij2024regulating,bonneau_sok_2015}.

\myparait{Placement.}
This initial stage introduces illicit funds into the financial system. Offenders often convert cash into legitimate-looking assets to avoid suspicion, using techniques like ``smurfing''---breaking large sums into smaller, less conspicuous transactions~\cite{sneck2024cryptocurrency,mooij2024regulating}. As the first point of contact with regulated systems, placement is both critical and risky~\cite{teichmann_recent_2020,mooij2024regulating}. In Web3, placement generally involves converting illicit fiat into crypto assets~(\S\ref{sec:background}), often via loosely regulated \textit{centralized exchanges}~(CEXs) to evade detection~\cite{sneck2024cryptocurrency,comolli_surfing_2021}.

\myparait{Layering.}
This stage follows placement and involves fabricating transaction records to conceal the origin of illicit funds~\cite{mooij2024regulating}, typically through a series of complex transactions designed to distance the assets from their source~\cite{sneck2024cryptocurrency}. In Web3, this often involves employing obfuscation strategies (discussed in~\S\ref{subsec:ml-strategies}) that reduce funds' traceability.

\myparait{Integration.}
This final stage reintroduces laundered funds into the legitimate economy, facilitating their use without arousing suspicion~\cite{sneck2024cryptocurrency}. This stage often entails the purchase of high-value assets such as real estate, luxury goods, or artwork~\cite{sneck2024cryptocurrency}. Within the Web3 domain, integration may also involve acquiring digital valuables, including NFTs~\cite{sneck2024cryptocurrency}. Moreover, some frameworks classify ``off-ramping''---the conversion of cryptocurrency into fiat and its reintegration into the traditional financial system---as a form of integration, particularly when disguised as capital gains or investment returns~\cite{gabbiadini2024money}.

While some scholars have noted limitations of the classical PLI framework in the context of emerging financial technologies (e.g., mapping specific behaviors to distinct PLI stages can be challenging with cryptocurrencies) ~\cite{pliIsOld2024,cassella2018toward}, it continues to serve as a foundational model in money laundering research~\cite{comolli_surfing_2021}. We thus structure our taxonomy on crypto-enabled money laundering around the PLI model, while accounting for domain‑specific particularities. For instance, in certain Web3-native crimes such as fraud, theft, or scams, the placement phase may be less distinct or entirely bypassed, as illicit proceeds originate within the crypto ecosystem~\cite{blockchaingroup2025,wu2023financial}. In such cases, laundering often begins with layering and proceeds to integration via off-ramping or asset conversion~\cite{chainalysis_report_2024}.

The next section examines the pseudonymous nature of the Web3 ecosystem and its role in laundering, outlines tiers of anonymity, and highlights a core challenge to deanonymization: limited access to data traces that hampers AML efforts.

\subsection{Deanonymization in Web3}
\label{subsection:anonimity-tiers}
Anonymity in cryptocurrencies is mainly assessed along two dimensions: preventing transaction \textit{linkage} and concealing user \textit{identities}~\cite{alsalami2019sok}. Alsalami et al.\cite{alsalami2019sok} outline four anonymity tiers: (a) \textit{Pseudonymity}, where users are identified by blockchain addresses that are not inherently tied to real-world identities; (b) \textit{Set anonymity}, which conceals a user's identity by embedding them within a broader group---called ``anonymity set''---such that the sender is indistinguishable from others in the set, typically through mechanisms like ring signatures (e.g., Monero~\cite{monero}); (c) \textit{Full anonymity}, which extends beyond set anonymity by cryptographically hiding both the sender's identity and transaction path, using techniques such as zero-knowledge proofs and commitment schemes (e.g., Zcash~\cite{zcash}); and (d) \textit{Confidential transactions}, which expands full anonymity by obscuring transaction amounts to ensure value privacy (e.g., Monero’s Bulletproofs).

Law enforcement has increasingly turned to blockchain forensics~\cite{deuber2022sok}, engaging firms such as Chainalysis~\cite{chainalysis} and Elliptic~\cite{elliptic} to assist in tracing cryptocurrency flows. Deanonymization often relies on correlating on-chain activity with off-chain data, particularly from gatekeepers such as CEXs that enforce \textit{know your customer}~(KYC) policies designed to associate customers with prosecutable real-world identities~\cite{comolli_surfing_2021}. When illicit assets interact with these services, investigators can link wallet addresses to real-world identities. On-chain graph analysis supports this process~\cite{alsalami2019sok,caringella2024bach}. Deanonymization and AML detection measures are often data-driven~(\S\ref{sec:ml-defenses}). Understanding data nature, scope, and accessibility is key for post-incident tracing and proactive measures. Next, we outline the typology of data used for data-driven money laundering countermeasures.

\subsection{Typologies of Data for Web3 AML Forensics}
\label{subsec:data-typology}

The data used for tracing illicit activity in Web3 can be broadly categorized into two types:

\mypara{On-chain data.}
This data type refers to information immutably recorded on public blockchains~\cite{comolli_surfing_2021,vaziry2024sok}, including transaction histories, addresses, amounts, fees, and timestamps~\cite{deuber2022sok}. Blockchains based on the \textit{unspent transaction output} (UTXO) model (e.g., Bitcoin~\cite{nakamoto2008bitcoin}) record UTXOs' ownership changes, while smart contract platforms (e.g., Ethereum~\cite{buterin2014next}) log state transitions, events, and token transfers~\cite{wernerSoKDecentralizedFinance2023,wood2025ethereum}.  On-chain data is publicly accessible through APIs or explorers like Etherscan~\cite{etherscan}, Bitquery~\cite{bitquery}, and The Graph~\cite{thegraph}. However, privacy-focused blockchains (e.g., Monero~\cite{monero}, Zcash~\cite{zcash}) deliberately restrict transparency~\cite{deuber2022sok}. 

\mypara{Off-chain data.}
This data type originates outside blockchains, and is stored by centralized entities~\cite{vaziry2024sok}. Sources include CEXs and custodial wallets, which often comply with KYC/AML regulations~\cite{comolli_surfing_2021}. These entities collect data such as names, IDs, IPs, device fingerprints, and activity logs~\cite{krakenPrivacy,chainalysis2022,blockchaingroup2025} that are accessible to investigators through legal processes. Additional sources include messaging platforms (e.g., Telegram, Discord), dark web listings, and communication logs~\cite{blockchaingroup2025,europol2017internet}, which provide context on laundering coordination. Analytics firms and regulators also maintain databases of flagged addresses and known scam networks~\cite{chainalysis2022}. While fragmented and variable in quality, off-chain data is essential for linking blockchain activity to real-world actors~\cite{mooij2024regulating}.

Notably, \textit{cross-chain transactional traces} are vital for detecting asset movement between blockchains—a process known as \textit{chain-hopping} (see~\S\ref{subsubsec:chain-hopping})\cite{lin2023towards}. These traces are not separate from on- or off-chain data but may consist of either—or both—depending on the method used. For example, on-chain smart contract logs from decentralized bridges (e.g., Wormhole~\cite{wormhole}, Multichain~\cite{multichain}) reflect token locking, minting, or burning~\cite{ou2022overview,gabbiadini2024money}, while atomic swaps using \textit{hashed timelock contracts} (HTLCs)~\cite{comolli_surfing_2021,gudgeon2019sok} or interoperability protocols like LayerZero~\cite{layerzero} and Axelar~\cite{axelar} produce traceable on-chain interactions~\cite{ou2022overview}. In contrast, chain-hopping via centralized services leaves only basic on-chain deposit and withdrawal traces, with most linkage data held off-chain in custodial databases~\cite{merklescience_chainhopping_2022}. Finally, some methods—such as those in \S\ref{subsubsec:disconnect-fund-flow}—generate no cross-chain traces at all.

Tracing illicit asset flows---especially across obfuscation layers---requires correlating on-chain and off-chain data from different sources. Each source offers a partial view; only by combining them can investigators fully reconstruct laundering paths. Sophisticated actors exploit gaps in data, jurisdiction, and privacy tools to evade detection~\cite{comolli_surfing_2021}.

\section{Taxonomy of Crypto Money Laundering}
\label{sec:ml-schemes}

This section presents a systematization of money laundering strategies on Web3 that apply across the PLI stages (\S\ref{subsec:ml-strategies}). Then, we examine intermediary technologies---ranging from privacy-enhancing tools to repurposed Web3 infrastructure---that facilitate asset obfuscation (\S\ref{subsec:ml-intermediaries}). Intermediaries often implement one or more of the above strategies (see Fig.~\ref{fig:taxonomy}).

\subsection{General Strategies}
\label{subsec:ml-strategies}

In cryptocurrency-based money laundering, most schemes rely on a core set of tactics to obscure the origin, flow, or ownership of illicit assets. We group these strategies into six broad categories. In practice, implementations often combine or blur these categories, making it difficult to assign a scheme to a single tactic. Still, our classification provides a systematic lens for understanding laundering patterns.

\subsubsection{Transaction Flow Obfuscation (TFO)}
\label{subsubsec:transaction-flow-obfuscation}

\begin{figure*}[t]
    \centering
    \includegraphics[width=0.99\textwidth]{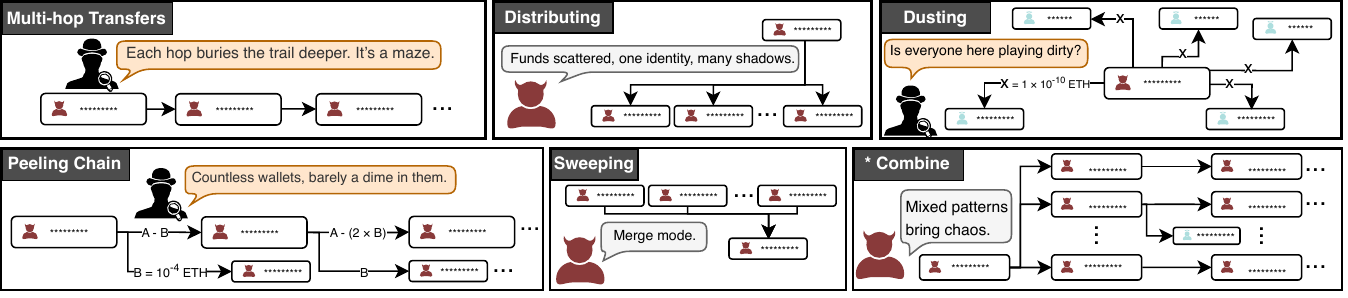}
    \vspace{-0.2cm}
    \caption[Transaction flow obfuscation (TFO) patterns.]{Transaction flow obfuscation (TFO) patterns.}
    \vspace{-0.3cm}
    \label{fig:TFO}
\end{figure*}

Transaction graph analysis as a means to detect illicit transactions is inherently error-prone, often yielding false positives due to the structural complexity of blockchain transactions~\cite{caringella2024bach,bonneau_sok_2015}. \textit{Transaction flow obfuscation} (TFO) refers to techniques that deliberately increase the complexity of blockchain transaction patterns, often by moving assets between wallets controlled by the same actor, mimicking usual transactional patterns~\cite{chang2018improving,shojaeinasab2023mixing}. As noted in~\S\ref{sec:background}, the ability of rapid, low-cost creation of new addresses is often exploited to obfuscate illicit fund flows~\cite{bonneau_sok_2015,mooij2024regulating}. Dispersing assets across many wallets increases transaction graph complexity and hinders forensic methods like address clustering~\cite{caringella2024bach,merklescience_hideTracks_nodate,lin2023towards}. Below, we outline common TFO patterns (see Fig.~\ref{fig:TFO}).

\mypara{Multi-hop transfers.}  
Chainalysis reports that over 80\% of illicit value flows involve routing funds through intermediary wallets---distinct addresses placed between the illicit source and the final conversion service---to increase obfuscation~\cite{chainalysis2024moneylaundering}. Each intermediary ``hop'' represents a transfer between addresses~\cite{malavolta2018anonymous,kolachala2021sok}. Multi-hop transfers often form long chains of ``relay transactions''~\cite{chang2018improving}, complicating manual tracing, e.g., when using public block explorers.

\mypara{Peeling chain.}
A \textit{peeling chain} is a transaction pattern commonly used by mixing services~(\S\ref{subsubsec:mixers}) to obfuscate the origin of large cryptocurrency holdings through incremental transfers~\cite{caringella2024bach}. It consists of a sequence of ``peel transactions''—each with one input and two outputs~\cite{chang2018improving,arbabi2023mixing}. Starting from a high-balance address, a small amount is sent to an external address while the remainder returns to a sender-controlled address~\cite{caringella2024bach}. This process recurses until the balance is fully dispersed, often ending with conversion at exchanges into fiat or other assets~\cite{caringella2024bach,merklescience_hideTracks_nodate,slowmist2023peelchain}. These patterns can generate false positives, as they resemble Bitcoin’s ``change address'' behavior. Under Bitcoin’s UTXO model, outputs must be fully spent, and excess funds are typically returned to a newly generated change address~\cite{caringella2024bach,chang2018improving}. This similarity complicates distinguishing illicit from legitimate transactions~\cite{caringella2024bach,chang2018improving}.

\mypara{Distributing and sweeping.}
``Distributing transactions'' typically involve any number of inputs and more than three outputs and are used to split large sums into smaller outputs~\cite{shojaeinasab2023mixing,chang2018improving}. They are often associated with a single entity sending funds to multiple recipients, as seen in gambling payouts or pooled mining rewards. These patterns, like peeling chains, relate to smurfing---the practice of breaking large amounts into smaller units to avoid detection~\cite{mooij2024regulating}. In contrast, ``sweep transactions'' consolidate multiple inputs into a single output~\cite{chang2018improving}. Commonly used by organizations for audit transparency or fund management, they are also employed to aggregate stolen assets from compromised wallets~\cite{shojaeinasab2023mixing,chang2018improving}.

\mypara{Dusting.}
Dusting term originally refers to the practice of sending negligible amounts of cryptocurrency---called ``dust''---to multiple wallets to trace and de-anonymize users by monitoring subsequent UTXO spending~\cite{wang2018anti,coinbase2025dusting,trezorDusting}. It is employed by malicious actors for extortion, by law enforcement for investigations, and by analytics firms for research~\cite{coinbase2025dusting}. In account-based systems like Ethereum, dusting has evolved into a phishing technique, where adversaries send counterfeit tokens embedded with malicious smart contracts. Interacting with these tokens---for example, through swaps or approvals---can lead to wallet compromise~\cite{trezorDusting}. Dusting is also used for ``address poisoning'', a phishing scheme in which attackers send small transactions from addresses resembling a victim’s address (e.g., with matching initial and final characters)~\cite{merklesciencePoisoning}. These addresses appear in transaction histories, increasing the chance of accidental reuse and misdirected funds.

Although not categorized as an explicit obfuscation attempt, in October 2018, the BestMixer.io mixer sent small Bitcoin amounts to numerous addresses---ostensibly for advertising~\cite{crawford2020knowing}. However, this act is widely interpreted as an attempt to disrupt AML systems by contaminating clean addresses, expanding the anonymity set and generating false positives in taint analysis~(see~\S\ref{subsec:6.2-preventive})~\cite{crawford2020knowing, alsalami2019sok}.

\subsubsection{Commingling}
\label{subsubsec:commingling}

In financial and legal contexts, \textit{commingling} denotes the intentional mixing of funds from multiple sources to obscure their origin and hinder traceability~\cite{cornell_commingling, comolli_surfing_2021, cassella2018toward}. It was central to 1990s warehouse banking schemes, where pooled client deposits masked individual ownership, facilitating tax evasion and money laundering~\cite{comolli_surfing_2021}. In Web3, commingling commonly occurs during layering, where assets from multiple parties are merged within a shared transactional or custodial structure. Unlike TFO~(\S\ref{subsubsec:transaction-flow-obfuscation}), where a single actor moves funds across self-controlled addresses or dusting recipients without their participation, commingling requires multiple parties. This shifts pseudonymity to set anonymity, thus probabilistically assigning fund origins to a group~\cite{alsalami2019sok}.

The terms \textit{commingling}, \textit{mixing}, \textit{tumbling}, and \textit{swapping} are closely related in cryptocurrency literature and often lead to confusion~\cite{comolli_surfing_2021,tairi20212}. To clarify, in this paper we use \textit{commingling} as an umbrella term referring to the pooling of assets from distinct parties to obscure their provenance~\cite{cornell_commingling}. \textit{Swapping}, by contrast, involves shuffling transaction destinations to achieve unlinkability~\cite{wu2021towards,shojaeinasab_decoding_2024,arbabi2023mixing} (\S\ref{subsubsec:disconnect-fund-flow}). \textit{Mixing} is often used as a broader term in the literature, encompassing various actions employed in \textit{mixers}~(\S\ref{subsubsec:mixers}), including TFO~(\S\ref{subsubsec:transaction-flow-obfuscation}), swapping~(\S\ref{subsubsec:disconnect-fund-flow}), and commingling~\cite{wu2021towards,comolli_surfing_2021}. Although the terms \textit{mixing} and \textit{tumbling} are often used interchangeably, some sources distinguish them. Tumbling refers to fragmenting transactions into smaller parts to obscure transaction trails~\cite{blockchaingroup2025,coinmarketcap_tumbler}, placing it under the TFO category and close to patterns such as peeling chains and distribution.

\subsubsection{Chain-hopping}
\label{subsubsec:chain-hopping}

Chain-hopping is an advanced obfuscation technique involving the deliberate transfer of assets across multiple blockchain networks to disrupt traceability~\cite{arbabi2023mixing,comolli_surfing_2021}. It poses significant challenges for investigators, who must aggregate and correlate transactional data from heterogeneous platforms---using cross-chain traces with on-chain or off-chain origins~(\S\ref{subsec:data-typology})---to attempt path reconstruction for inferring ownership and track illicit flows~\cite{kolachala2021sok}. Anonymity-enhanced blockchains (\S\ref{subsubsec:AECs}) further complicate forensic analysis~\cite{comolli_surfing_2021}. 

Chain-hopping may occur via CEXs, where custodial platforms convert assets across chains~\cite{almeida2023review,comolli_surfing_2021}, or via DEXs using liquidity pools and atomic swaps~\cite{herlihy2018atomic}, enabling peer-to-peer cross-chain transfers without intermediaries~\cite{wernerSoKDecentralizedFinance2023,comolli_surfing_2021}(\S\ref{subsubsec:exchanges}). Cross-chain bridges\cite{hu2024piecing}, which lock assets on one chain and release equivalents on another, are also key enablers of chain-hopping~(\S\ref{subsubsec:bridges}). All such chain-hopping enablers are detailed in \S\ref{subsec:ml-intermediaries}.

\subsubsection{Disconnected fund flow (DFF)}
\label{subsubsec:disconnect-fund-flow}

While previously discussed tactics often leave residual data traces that can support partial linkage between source and destination account, a distinct class of techniques---referred to as \textit{disconnected fund flow} (DFF)~\cite{arbabi2023mixing,shojaeinasab_decoding_2024}---aims to eliminate any traceable connection between illicit origins and final destinations. And while earlier tactics typically offer set anonymity~\cite{alsalami2019sok}---providing limited resistance to deanonymization---DFF can approach full anonymity~\cite{alsalami2019sok}~(see \S\ref{subsection:anonimity-tiers}). Finally, unlike prior layering-specific strategies, DFF techniques are also applicable during placement and integration. Below, we explore these tactics.

\begin{figure*}[t]
    \centering
    \includegraphics[width=0.99\textwidth]{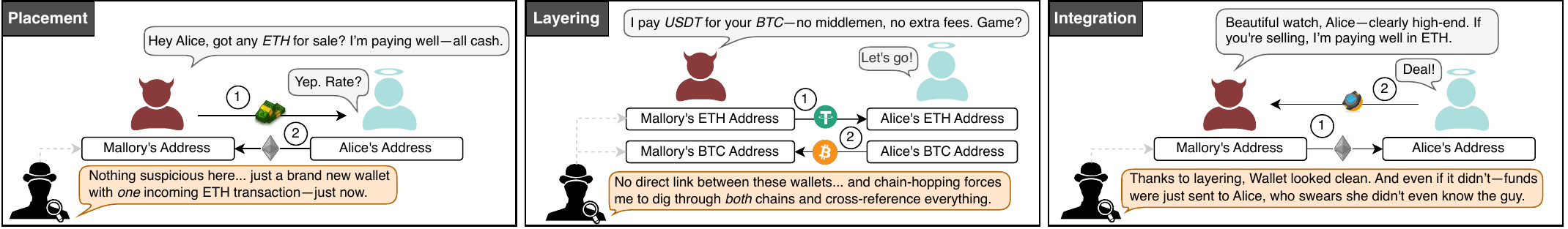}
    \vspace{-0.2cm}
    \caption[Disjoint reciprocal transfers (DRT).]{Disjoint reciprocal transfers (DRT).}
    \vspace{-0.3cm}
    \label{fig:DRT}
\end{figure*}

\mypara{Disjoint reciprocal transfers (DRTs).} 
A representative example of DFF is a process we refer to as \textit{disjoint reciprocal transfers}~(DRTs), after its structural characteristics---two reciprocal P2P transactions intentionally designed to appear unrelated~(see Fig.~\ref{fig:DRT}). These transactions lack observable links, rendering their connection untraceable or extremely difficult to reconstruct, even with advanced forensic tools~\cite{ruffing2017p2p,almeida2023review,comolli_surfing_2021}.
%\begin{itemize}

\myparait{Placement via DRTs.}  
In the placement phase, launderers introduce illicit funds---often in the form of physical cash---into the cryptocurrency ecosystem via P2P asset swaps with third parties~\cite{almeida2023review,comolli_surfing_2021}. These exchanges typically occur through informal channels such as online forums, encrypted messaging apps, or social media. The actor transfers tainted funds to a counterparty, who returns an equivalent amount of clean cryptocurrency. 
With minimal contact through safe communication channels, these exchanges must be protected against deanonymization. Given the placement stage’s inherent risk~\cite{teichmann_recent_2020,mooij2024regulating,sneck2024cryptocurrency}, launderers exercise caution by avoiding traceable interactions and leverage \textit{privacy-enhancing technologies}~(PETs) such as \textit{Tor}\cite{dingledine2004tor} to obscure personal identifiers and conceal network-level metadata\cite{moser_inquiry_2013}.

\myparait{Layering via DRTs.}  
Layering can leverage DRTs, enabling a malicious actor (Mallory) to exchange tainted cryptocurrency for ostensibly clean assets from a third party (Alice) without leaving discernible links. When both transaction happen on the same blockchain, both parties must use distinct addresses for sending and receiving (Mallory$_0$~$\rightarrow$~Alice$_0$, Alice$_1$~$\rightarrow$~Mallory$_1$), since reusing any address exposes linkages. Related work on stealth address schemes similarly seeks to preserve recipient anonymity by rendering repeated payments unlinkable~\cite{courtois2017stealth}. Furthermore, DRTs can facilitate chain-hopping~(\S\ref{subsubsec:chain-hopping}). While requesting address changes may raise suspicion in single-chain transfers, it appears natural when assets are exchanged across blockchains---e.g., Bitcoin for USDT. As discussed in~\S\ref{sec:background}, addresses on different blockchains, particularly those derived from hierarchical deterministic (HD) wallets~\cite{bip44}, are structurally unrelated. This address discontinuity breaks implicit links.

\myparait{Integration via DRTs.}  
DRTs can also serve as conduits for reintroducing laundered assets into the legitimate economy~\cite{almeida2023review}. Adversaries may employ layered crypto assets to acquire high-value goods—such as real estate, vehicles, luxury commodities, or high-end NFTs—through transactions with legitimate third parties, thereby completing the laundering cycle.

Related to DRTs are over-the-counter (OTC) and P2P trades~\cite{comolli_surfing_2021,sneck2024cryptocurrency}, often facilitated by CEXs, such as KuCoin’s P2P interface for USD–USDT swaps~\cite{kucoinP2POTC}. Here, the platform acts as an intermediary, connecting laundering actors with third parties. However, CEXs involvement introduces a degree of transparency that can undermine the disjoint nature of the transaction. While CEXs retain transactional records and are subject to KYC and AML regulations~\cite{comolli_surfing_2021}, malicious actors often exploit regulatory gaps to avoid identification (\S\ref{subsubsec:exploit-regulatory-gaps}).

\mypara{Swapping.}  
Swapping is also a DFF technique that shuffles transaction destinations to achieve unlinkability by decoupling user inputs and outputs, thereby preserving address anonymity~\cite{wu2021towards,shojaeinasab_decoding_2024,arbabi2023mixing}. For instance, a swap can map transactions as follows:

{\small
\[
\text{(Intended)}\hspace{5pt}
\text{A} \xrightarrow{20} \text{D},\ 
\text{B} \xrightarrow{10} \text{E},\ 
\text{C} \xrightarrow{10} \text{F}
\hspace{5pt}\]
\[
\text{(Final)}\hspace{5pt}
\text{A} \xrightarrow{10} \text{E},\ 
\text{A} \xrightarrow{10} \text{F},\ 
\text{B} \xrightarrow{10} \text{D},\ 
\text{C} \xrightarrow{10} \text{D}
\]}

\mypara{Zero-knowledge proofs (ZKPs).}  
ZKPs are protocols that let one party prove a statement’s validity to another without revealing underlying information~\cite{goldwasser1985knowledge,fiege1987zero}. Privacy-focused cryptocurrencies (e.g., Zerocash~\cite{sasson2014zerocash}) adopted ZKPs to balance transparency, security, and efficiency~\cite{comolli_surfing_2021, arbabi2023mixing, shojaeinasab_decoding_2024}. ZKPs can provide transaction unlinkability and full anonymity~\cite{alsalami2019sok}, and are used in PETs across the crypto ecosystem and Web3 and DeFi infrastructures (see~\S\ref{subsec:ml-intermediaries}).

\subsubsection{Exploit Subjective Valuations}
\label{subsubsec:exploit-subjective-valuation}

Consider colluding malicious actors, \textit{Mallory} and \textit{Darth}. To obscure a \$50,000 illicit payment, Mallory initially acquires a low-value artwork (e.g., \$50) and orchestrates its resale through collusive bidding to artificially inflate its price. Darth subsequently purchases the piece for \$50,050, legitimizing the transaction by attributing the overpayment to sentimental value. Mallory, in turn, justifies the windfall as a capital gain, thereby laundering the funds under the guise of a legitimate art sale. The subjective nature of art valuation shields both parties from direct legal scrutiny~\cite{financialcrimeacademy_art_2023,schneider2004money}. FATF’s 2023 report underscores how the art market’s opacity and lax regulation enable such exploitation~\cite{FATF2023art}. Unlike fungible assets (e.g., fiat, gold), which are uniformly priced, non-fungible assets are unique, with volatile and opaque valuations. NFTs~(\S\ref{subsubsec:nfts}) brought subjective valuation into the digital realm, enabling laundering~\cite{mooij2024regulating,mekacher2022heterogeneous,ziemke2023determines}. A prominent example is \textit{The Merge}, which sold for \$91.8 million~\cite{wang2023nfts}.

\subsubsection{Exploit Regulatory Gaps}
\label{subsubsec:exploit-regulatory-gaps}

Exploiting regulatory gaps is a core strategy in crypto money laundering~\cite{campbell2018bitcoin}. Launderers leverage inconsistent AML enforcement across jurisdictions, exploiting regulatory fragmentation to move illicit funds~\cite{comolli_surfing_2021,campbell2018bitcoin}.
KYC regulations are central to AML compliance, requiring service providers and exchanges to verify identities, monitor transactions, and report suspicious activity~\cite{comolli_surfing_2021,mooij2024regulating,slowmist2024report}. However, many offshore or non-compliant cryptocurrency services lack KYC controls, enabling anonymous transactions and regulatory evasion~\cite{comolli_surfing_2021}. \textit{Regulatory arbitrage} is a tactic where launderers shift operations to jurisdictions with lax AML oversight, exploiting systemic gaps and hindering international enforcement~\cite{comolli_surfing_2021,slowmist2024report}. The global expansion of decentralized financial ecosystems and VASPs in regions with inconsistent regulation, such as parts of Asia-Pacific and Latin America, amplifies these risks~\cite{slowmist2024report}.

\subsection{Intermediary Mechanisms}
\label{subsec:ml-intermediaries}

While general laundering strategies in the Web3 ecosystem~(\S\ref{subsec:ml-strategies}) address high-level tactics for obscuring the provenance of illicit assets, a critical enabler of these strategies lies in the exploitation of specific intermediary mechanisms. These intermediaries---including PETs as well as repurposed decentralized infrastructures and financial instruments---function as infrastructural conduits that facilitate the placement, layering, and integration stages of money laundering. In this section, we systematically examine the principal intermediary mechanisms exploited in crypto money laundering, analyzing how they are either purpose-built or opportunistically repurposed to enhance obfuscation, frustrate traceability, and evade regulatory scrutiny across diverse laundering schemes.

\subsubsection{Mixing Services}
\label{subsubsec:mixers}

\textit{Mixing services}---also known as \textit{tumblers}, \textit{cryptocurrency mixers}, or \textit{coin mixing services}---are the primary transaction anonymity tools in Web3. They have been widely studied across business, finance, computer science, and legal fields~\cite{glaeser2022foundations, crawford2020knowing, comolli_surfing_2021, campbell2018bitcoin, ruffing2017p2p, lin2023towards}. 

Mixers emerged shortly after Bitcoin, reflecting early concerns over transaction privacy~\cite{moser_inquiry_2013}. Services like BitLaundry~\cite{bitlaundry} and Bitcoin Fog~\cite{bitcoinfog} pooled and redistributed funds to conceal provenance. Bitcoin's ties to illicit marketplaces boosted demand for mixers~\cite{bonneau_sok_2015}, which became key to laundering darknet and ransomware proceeds~\cite{hong2018poster}. From 2015 to 2017, mixer transaction volumes quadrupled~\cite{hong2018poster}, highlighting their role in the crypto economy.

\myparait{Layering via Mixers.}
Mixers commonly employ TFO patterns (\S\ref{subsubsec:transaction-flow-obfuscation}), distorting transaction paths and hinder taint analysis~\cite{shojaeinasab_decoding_2024,caringella2024bach}. Notably, \textit{tumbling}---breaking funds into smaller sums to obscure the transaction trail---is a canonical TFO pattern. This is why mixers are often referred to as tumblers~\cite{blockchaingroup2025, coinmarketcap_tumbler}. Historically, commingling was also a foundational concept of mixers~\cite{comolli_surfing_2021}. Early implementations exploited Bitcoin's UTXO model to sever address ownership linkage~\cite{bonneau_sok_2015}.
CoinJoin~\cite{coinjoin}, for instance, allowed multiple users to aggregate UTXOs into a single transaction, making input-output associations difficult. Modern mixers retain this principle, pooling transactions from multiple parties and redistributing them to obscure traceability~\cite{shojaeinasab_decoding_2024, moser_inquiry_2013}. As discussed in~\S\ref{subsection:anonimity-tiers}, mixers enhance \textit{set anonymity}~\cite{alsalami2019sok}, complicating the tracing of funds back to true sources~\cite{shojaeinasab_decoding_2024}.

Modern mixers also utilize advanced DFF techniques (\S\ref{subsubsec:disconnect-fund-flow}), such as ZKPs, to obscure fund flows, sever sender-receiver linkages, and hinder taint analysis~\cite{shojaeinasab_decoding_2024}. As noted by Wu et al.~\cite{wu2021towards}, mixers generally follow two paradigms: \textit{obfuscation}~(see \S\ref{subsubsec:transaction-flow-obfuscation}) and \textit{swapping}~(see \S~\ref{subsubsec:disconnect-fund-flow}).

Mixing across blockchain platforms enables chain-hopping. Services like ShapeShift enabled crypto swaps without custodial risk~\cite{wu2021towards, arbabi2023mixing}, but also facilitated chain-hopping. A Wall Street Journal report linked ShapeShift to at least \$9 million in laundered assets~\cite{arbabi2023mixing}.
 
Mixing services typically charge transaction fees of 0.5\%–4\%~\cite{hong2018poster}. While these fees constitute operator revenue, competitive pricing can also expand the anonymity set and strengthen mixing effectiveness. Early mixers relied on fixed fees and deterministic delays, which enabled de-mixing attacks; in contrast, modern mixers employ randomized fees, variable delays, and staggered outputs to disrupt input–output heuristics and timing correlations~\cite{shojaeinasab_decoding_2024}. Early mixers employed custodial aggregation, recording transactions off-chain while only deposits and withdrawals appeared on-chain~\cite{wu2021towards}. Trust concerns, including exit scams and regulatory risk, drove the shift to decentralized, smart contract–based protocols such as \textit{Tornado Cash}, which operate non-custodial~\cite{glaeser2022foundations,shojaeinasab_decoding_2024,lin2023towards}.

\subsubsection{Anonymity-Enhanced Cryptocurrencies (AECs)}
\label{subsubsec:AECs}

Some blockchains are designed to obscure transaction details, including sender/receiver pairs and amounts~\cite{comolli_surfing_2021,alsalami2019sok}. AECs—also known as ``privacy coins''—employ advanced privacy technologies that go beyond standard pseudonymity, offering stronger anonymity. This impedes forensic analysis and can be exploited to facilitate money laundering~\cite{campbell2018bitcoin}.

\myparait{Layering via AECs.} AECs support the \textit{layering} stage by severing links between illicit funds and their origins~\cite{comolli_surfing_2021}, using various privacy techniques. Monero~\cite{monero} employs \textit{ring signatures}, \textit{stealth addresses}, and confidential transactions to anonymize identities and amounts~\cite{alsalami2019sok}, mixing real inputs with decoys to implement \textit{DFF}(\S\ref{subsubsec:disconnect-fund-flow}). Zerocash\cite{sasson2014zerocash}, implemented in Zcash~\cite{zcash}, uses \textit{zk-SNARKs} for fully anonymous transactions; Zcash supports both \textit{transparent} and \textit{shielded} modes, the latter hiding inputs, outputs, and amounts~\cite{alsalami2019sok}, thereby achieving DFF. In contrast, Dash~\cite{dash} applies CoinJoin to merge multiple users’ transactions, obscuring individual trails by commingling~(\S\ref{subsubsec:commingling})~\cite{bonneau_sok_2015}.

\subsubsection{Cryptocurrency Exchanges}
\label{subsubsec:exchanges}

Cryptocurrency exchanges are pivotal in the Web3 ecosystem, enabling conversions between fiat and various crypto assets~\cite{comolli_surfing_2021}. Their central role makes them exploitable across all money laundering stages: placement, layering, and integration~\cite{comolli_surfing_2021, mooij2024regulating}. 

\mypara{Centralized exchanges (CEXs).} 
CEXs act as intermediaries for asset trading~\cite{comolli_surfing_2021}, managing users' assets via internal custodial wallets~\cite{mooij2024regulating} funded with fiat or cryptocurrencies. 

\myparait{Placement via CEXs.}
As crypto ecosystem gatekeepers, CEXs are frequently exploited for placement in weak regulation settings~(\S\ref{subsubsec:exploit-regulatory-gaps})~\cite{comolli_surfing_2021}. Criminals leverage offshore and no-KYC exchanges to convert illicit fiat into crypto, often using smurfs or money mules---multiple accounts depositing smaller amounts to avoid suspicion~\cite{comolli_surfing_2021,gabbiadini2024money}. Integrated P2P and OTC services within CEXs~\cite{kucoinP2POTC} facilitate DRTs, enabling fiat-to-crypto conversions(\S\ref{subsubsec:disconnect-fund-flow}) that disconnects fund flows by exploiting regulatory gaps~\cite{gabbiadini2024money,chainalysis2024moneylaundering}.

\myparait{Layering via CEXs.}
Many CEXs support off-chain internal transfers between user accounts~\cite{bybit_internal_transfer,kucoin_internal_transfer,binance_internal_transfer,bitfinex_internal_transfer,coinbase_internal_transfer}, typically free, functioning as simple database operations. These enable off-chain fund movements, with no on-chain traces. In lax compliance settings where KYC measures are absent, such movements evade both on-chain and off-chain tracing, contributing to DFF~(\S\ref{subsubsec:disconnect-fund-flow}). CEXs also facilitate \textit{chain-hopping}~(\S\ref{subsubsec:chain-hopping}) by allowing users to deposit assets on one blockchain and withdraw on another~\cite{gabbiadini2024money}. Finally, while most reputable exchanges now assign unique deposit addresses to each user powered by HD wallets, there have been instances where exchanges have reused addresses across multiple users~\cite{victor2020address,stackexchange_address_reuse}. This contributes to obfuscating transaction flows~(\S\ref{subsubsec:transaction-flow-obfuscation}) when conducting taint analysis~\cite{victor2020address}, as well as aggregating funds from multiple users in one place, inherently contributing to \textit{commingling}~(\S\ref{subsubsec:commingling}).

\myparait{Integration via CEXs.}
CEXs may serve as off-ramps for converting laundered crypto-assets into fiat, reintroducing them into the traditional economy~\cite{mooij2024regulating,chainalysis2024moneylaundering,lin2023towards}. In jurisdictions with KYC/compliance gaps, OTC brokers offer launderers low-risk fiat exit points~\cite{chainalysis2024moneylaundering}.

%\end{itemize}

\mypara{Decentralized exchanges (DEXs).} 
DEXs operate autonomously via smart contracts, without centralized intermediaries~\cite{xu2023sok,comolli_surfing_2021}. They automatically match users trading virtual assets. Beyond individual DEXs, aggregators synchronize with multiple platforms, pulling order book data to automate trades and offer best pricing~\cite{comolli_surfing_2021}. As DEXs do not interface with fiat, their laundering utility is mostly limited to layering.

%\begin{itemize}
    \myparait{Layering via DEXs.}
    Launderers employ \textit{TFO} by exchanging tainted assets across liquidity pools~\cite{lin2023towards,USTreasury2023DeFiRisk}. Atomic cross-chain swaps~\cite{herlihy2018atomic, malavolta2018anonymous} enable \textit{chain-hopping}~(\S\ref{subsubsec:chain-hopping}), allowing assets to traverse multiple blockchains with minimal traceability~\cite{USTreasury2023DeFiRisk,gabbiadini2024money,comolli_surfing_2021}.
%\end{itemize}

\subsubsection{Cross-chain Bridges}
\label{subsubsec:bridges}

These bridges enable interoperability between isolated blockchains, transferring assets, data, and messages~\cite{hu2024piecing, zhang2024security}. They typically use (a) \textit{lock-and-mint/burn-and-release} or (b) \textit{liquidity-pool-based models} to manage cross-chain flows~\cite{lee2023sok, zhang2024security}. In the former, tokens are locked (reversible) or burned (irreversible) on the source chain, while equivalent wrapped tokens (see~\S\ref{sec:background}) are minted on the destination~\cite{lee2023sok}. Liquidity-pool-based bridges bypass wrapped tokens, executing swaps using pools of original assets.

Bridging also applies to non-fungible tokens~\cite{harris2023cross}. \textit{Bridged NFTs} are NFTs transferred across chains via a similar lock-and-mint logic as described above.

%\begin{itemize}
\myparait{Layering via bridges.} Bridges enable \textit{layering} by facilitating chain-hopping~(\S\ref{subsubsec:chain-hopping})~\cite{lin2023towards,USTreasury2023DeFiRisk}, allowing launderers to move assets across blockchains and sever direct provenance links~\cite{hu2024piecing}, thus making it more difficult to trace these assets~\cite{zhang2024security, muhammad2024cross}. Past studies reveal that a substantial portion of cross-chain fund flows reaches exchanges, highlighting the operational integration of bridge-based chain-hopping into broader laundering pipelines~\cite{hu2024piecing}. Since NFT bridges facilitate the transfer of NFTs between blockchain networks~\cite{harris2023cross}, they might also be used to facilitate chain-hopping~(\S\ref{subsubsec:chain-hopping}).
%\end{itemize}

\subsubsection{DeFi Mechanisms}
\label{subsubsec:defi}

DeFi comprises primitives such as lending platforms, yield aggregators, and stablecoins, many of which can be exploited to add a extra layer of obfuscation in transaction flows~\cite{comolli_surfing_2021,wernerSoKDecentralizedFinance2023}. Although DeFi's decentralized nature and lack of fiat interaction limit its role in placement and integration, custodial stablecoin issuers extend DeFi's reach to these stages~\cite{moin2020sok}.

\mypara{Loan providers.}
Decentralized lending protocols facilitate permissionless borrowing through two primary mechanisms: over-collateralized loans and flash loans~\cite{wernerSoKDecentralizedFinance2023,gudgeon2020defi}. Over-collateralized loans resemble traditional debt arrangements, requiring users to lock assets exceeding the loan value. In contrast, flash loans allow users to borrow assets without collateral, under the condition that the borrowed amount is repaid within the same blockchain transaction. If the repayment condition is not met, the transaction is reverted.%, ensuring no risk to the lender.
%\begin{itemize}

    \myparait{Layering via over-collateralized loans.}
    Illicit assets can be pledged as collateral to obtain clean loaned funds~\cite{wernerSoKDecentralizedFinance2023}. Furthermore, by extending loans to third parties and receiving repayments, launderers facilitate commingling~(\S\ref{subsubsec:commingling}) of illicit and legitimate assets. Cross-chain lendings further support obfuscation via chain-hopping~(\S\ref{subsubsec:chain-hopping})~\cite{zhang2024security}.
    \myparait{Layering via flash loans.}
     Launderers can acquire flash loans and execute complex transactions within a single block, including wash trading---fake trades used as cover---and cycling assets before repaying the loan~\cite{gudgeon2020defi,comolli_surfing_2021}. Flash loans can also facilitate collateral swapping or the repayment of loans with different assets. These activities can result in TFO~(\S\ref{subsubsec:transaction-flow-obfuscation}) and the commingling of illicit funds with loaned assets~(\S\ref{subsubsec:commingling}).
%\end{itemize}

\mypara{Yield farming and liquidity pools.}
Yield farming refers to the practice of providing capital to DeFi protocols in exchange for financial rewards, typically in the form of governance or utility tokens~\cite{xu2023sok, wernerSoKDecentralizedFinance2023}. A key mechanism for this is liquidity provisioning to Automated Market Makers (AMMs), which replace traditional order books by enabling token swaps in DEXs through algorithmically managed liquidity pools~\cite{cousaert2022sok}. Users who supply assets to these pools earn a share of the transaction fees and may also receive protocol incentives. Yield farming can occur in both decentralized and custodial manners (eg., through CEXs).

%\begin{itemize}
    \myparait{Layering via yield farming.}
    Launderers invest tainted assets into yield farming to acquire rewards,  commingling~(\S\ref{subsubsec:commingling}) illicit funds with clean proceeds~\cite{lin2023towards,USTreasury2023DeFiRisk}. Rewards may be withdrawn on other blockchains---when using CEXs---facilitating chain-hopping~(\S\ref{subsubsec:chain-hopping}).

    \myparait{Layering via liquidity pools.}
    Launderers begin by creating fake tokens~\cite{lin2023towards}, which they pair with legitimate assets---such as ETH---within DEXs. This pairing involves supplying both assets to a new liquidity pool, enabling the counterfeit token to appear tradeable. By conducting wash trades between accounts they control, launderers simulate organic market activity and inflate the token's perceived legitimacy. This tactic commingles~(\S\ref{subsubsec:commingling}) illicit funds with on-chain liquidity while masking their origin. Ultimately, launderers extract clean ETH from the pool, disguised as ordinary DEX trading.
%\end{itemize}

\mypara{Stablecoins.}
Stablecoins---especially fiat-backed variants---serve as gateways between fiat economies and DeFi~\cite{kwon2023drives}. Custodial issuers (e.g., Circle~\cite{USDC}, Tether~\cite{USDT}) enable the conversion of illicit assets into legitimate digital dollars, facilitating placement and integration when KYC measures are circumvented. In contrast, crypto-collateralized stablecoins like Dai~\cite{Dai}, which operate primarily as over-collateralized loan issuers, allow criminals to pledge illicit assets as collateral to mint new tokens---a process exploitable during layering~\cite{comolli_surfing_2021}.

\subsubsection{Non-Fungible Tokens (NFTs)}
\label{subsubsec:nfts}

Non-fungible tokens were first implemented on Ethereum via EIP-721~\cite{erc721} and later became available on Bitcoin through ``Ordinals,'' which inscribe arbitrary data onto serialized satoshis~\cite{ordinals}. In Web3, NFTs serve as digital ownership certificates, enabling verifiable scarcity~\cite{mooij2024regulating, hammi2023non}. NFTs are tightly coupled with Web3 metaverses. The term ``metaverse,'' coined in \textit{Snow Crash}~\cite{snowcrash}, has evolved into a vision of immersive, networked 3D environments. Inspired by MMO games~\cite{mooij2024regulating}, Web3 metaverses include platforms such as Decentraland~\cite{decentraland2024whitepaper} and The Sandbox~\cite{sandbox_whitepaper}, where NFTs represent avatars, skins, and virtual land parcels. Here, NFTs underpin digital property rights and enable economic transactions~\cite{mooij2024regulating}---acquiring metaverse properties often requires only a wallet, bypassing KYC checks~\cite{wu2023financial}.

NFTs have been examined as laundering vectors across all PLI stages~\cite{sneck2024cryptocurrency, mooij2024regulating, wu2023financial}. Metaverse platforms---where NFT trading is prevalent---offer fertile ground for laundering due to liquidity and ease of transfer of digital collectibles~\cite{mooij2024regulating}.

    \myparait{Placement via NFTs.}
    As described in~\S\ref{subsubsec:disconnect-fund-flow}, illicit fiat can be funneled into NFTs via disjoint reciprocal transfers, introducing tainted funds into the crypto ecosystem.

    \myparait{Layering via NFTs.}
    As discussed in~\S\ref{subsubsec:exploit-subjective-valuation}, NFT trading with subjective pricing contributes to layering, while bridged NFTs also facilitate chain-hopping~(\S\ref{subsubsec:bridges}).

    \myparait{Integration via NFTs.}
    As discussed in~\S\ref{sec:ml-primer} and~\S\ref{subsubsec:disconnect-fund-flow}, NFTs can act as value stores or speculative assets~\cite{sneck2024cryptocurrency}, and be sold to reintroduce cleaned funds into the legal economy.

\subsubsection{Cryptocurrency ATMs and Cards}
\label{subsubsec:atms-and-cards}

Here, we explore the role of ATMs and card-based systems in money laundering.

\mypara{ATMs.}
Cryptocurrency ATMs---also known as ``Bitcoin ATMs (BTMs)''~\cite{noll2023controversial}---enable access to crypto without online exchanges and can be found in stores or gas stations~\cite{comolli_surfing_2021,noll2023controversial}. To deposit cash and purchase crypto, users provide minimal ID (e.g., a phone number) and scan a receiver address via QR code~\cite{comolli_surfing_2021,noll2023controversial}; some BTMs generate paper wallets if no address is given~\cite{noll2023controversial}. For withdrawals, crypto is sent to a designated address and cash is dispensed.

BTMs have drawn scrutiny for illicit use~\cite{comolli_surfing_2021,sneck2024cryptocurrency}. TRM Labs reported that the cash-to-crypto sector---largely driven by BTMs---processed at least \$160M in illicit funds since 2019~\cite{trmlabs-illicit-atm}. BTMs are frequently used by dark web vendors to cash out illicit proceeds~\cite{comolli_surfing_2021}, prompting regulatory concern~\cite{trmlabs-illicit-atm}. Despite existing safeguards, inconsistent regulations and weak KYC measures allow launderers to exploit BTMs to inject illicit funds into the financial system or retrieve them as seemingly legitimate assets~\cite{comolli_surfing_2021,sneck2024cryptocurrency}.

\mypara{Cards.}
Several card-related mechanisms can be exploited for crypto money laundering:

\myparait{Integration via crypto-linked cards.}
Crypto-linked cards (e.g., Visa~\cite{visa2022crypto}, Mastercard~\cite{mastercardCryptoCard}) enable users to spend crypto directly, with issuers converting it to fiat at purchase time so merchants receive traditional currency~\cite{comolli_surfing_2021}. These cards---also offered by exchanges and payment processors---function like fiat-funded debit cards, enabling online/in-person payments and ATM withdrawals~\cite{visa-crypto-attitudes,osc-crypto-survey}. However, they pose laundering risks akin to traditional financial tools~\cite{comolli_surfing_2021}, enabling integration of laundered cryptocurrencies.

\myparait{Integration via gift-card shops.} Platforms like Bitrefill~\cite{bitrefill-getting-started} sell gift cards accepting crypto payments, often without strong KYC policies. During the integration phase, launderers can convert crypto into gift cards and resell them to third parties.

\myparait{Placement via pre-funded crypto gift-cards/vouchers.}
Sold in fiat, these products are pre-loaded with fixed crypto amounts and redeemed via embedded private keys or claim codes. Acquirable in bulk, they can be exploited for illicit fiat placement, particularly in weak-KYC jurisdictions~(\S\ref{subsubsec:exploit-regulatory-gaps}). \textit{Ballet} steel cards~\cite{bitdegreeBalletWalletReview,balletCryptoEasy}, \textit{Tangem Notes}~\cite{tangem,tangemSmartBanknotes}, \textit{Casascius}/\textit{Denarium} coins~\cite{casascius,denarium}, and digital vouchers from \textit{CryptoVoucher}, \textit{Bitnovo}, and \textit{Azteco}~\cite{cryptovoucher,bitnovo,dundleBitnovoVoucher,azteco} are notable examples. Although many providers have discontinued pre-loaded options—likely in response to rising regulatory scrutiny—such mechanisms remain accessible in some markets and are considered high-risk instruments~\cite{fintrac2024gambling}.

\subsubsection{Online Gambling Platforms}
\label{subsubsection:gambling}

Online gambling platforms are key vectors for financial obfuscation and money laundering~\cite{ sneck2024cryptocurrency,schneider2004money}. 
While most accept fiat and crypto, many crypto-native implementations operate with minimal regulation~\cite{sneck2024cryptocurrency}, exposing laundering vectors across all PLI stages~\cite{interpol-gambling-aml,fintrac2024gambling}. Smart contract-based gambling dApps further automate settlement and bypass AML controls~\cite{chainalysis-crypto-crime}. 

\myparait{Placement via gambling deposits.}
Illicit fiat or crypto is deposited into gambling accounts as wagers~\cite{sneck2024cryptocurrency}.

\myparait{Layering via gambling activity.}
Fragmented bets and intra-platform transfers create transaction graphs that mimic legitimate behavior~(\S\ref{subsubsec:transaction-flow-obfuscation})~\cite{almeida2023review,sneck2024cryptocurrency}. Illicit funds mix with real winnings~(\S\ref{subsubsec:commingling})~\cite{sneck2024cryptocurrency}. One laundering method is ``chip dumping,'' where a player intentionally loses PvP games (e.g., poker) to transfer funds to a co-conspirator~\cite{fintrac2024gambling}.

\myparait{Integration via withdrawals.}
Remaining balances, now masked as gambling proceeds, are withdrawn to wallets or bank accounts and misrepresented as legitimate income~\cite{comolli_surfing_2021}.

\subsubsection{Initial Coin Offerings (ICOs)}
\label{subsubsec:icos}

ICOs are blockchain-based fundraising mechanisms where ventures issue digital tokens in exchange for crypto or fiat~\cite{delivorias2021understanding, barone2019cryptocurrency}. Unlike traditional channels, they use white papers instead of regulated prospectuses and rely on online promotion rather than underwriting, offering global investor access with minimal legal barriers~\cite{delivorias2021understanding, barone2019cryptocurrency}. Issued tokens generally fall into three categories: currency (payment), utility (platform access), and investment (profit participation), though many are hybrids~\cite{klohn2018initial}. The absence of shareholder rights and governance structures deprives token holders of oversight and heightens fraud risks~\cite{delivorias2021understanding}. The decentralized nature and regulatory gaps~(\S\ref{subsubsec:exploit-regulatory-gaps}) make ICOs exploitable across PLI stages:
%\begin{itemize}

    \myparait{Placement via ICOs.}  
    Placement is feasible when ICOs accept fiat. Criminals can convert illicit cash into digital tokens. Weak KYC during early ICO phases~\cite{delivorias2021understanding} facilitated such entry, though fiat acceptance remains less common than crypto.

    \myparait{Layering via ICOs.}  
    More commonly, ICOs accept cryptocurrencies like BTC or ETH. In that case, illegal crypto funds are invested into ICOs, where they appear as legitimate contributions to a cryptocurrency project~\cite{barone2019cryptocurrency}. If the project succeeds, the funds are returned as profits, now disguised as lawful investment gains. Essentially, malicious actors invest tainted funds---commingling them with clean assets of other participants~(\S\ref{subsubsec:commingling})---and receive new tokens~(\S\ref{subsubsec:transaction-flow-obfuscation}). If the ICO sends tokens to a different address than the payer’s, it introduces DFF~(\S\ref{subsubsec:disconnect-fund-flow}). Accepting payments on one network (eg., Bitcoin) but issuing tokens on another (e.g., Ethereum) also enables \textit{chain-hopping}~(\S\ref{subsubsec:chain-hopping}).
    A more advanced mirroring tactic (\S\ref{subsubsec:defi}) is to create a counterfeit ICO, where actor $A$ hosts the ICO, and $B$ buys worthless tokens using illicit crypto. The transaction appears as a speculative loss but facilitates value transfer, exploiting justifiable price volatility.

    \myparait{Integration via ICOs.}  
    If a counterfeit ICO accepts fiat, the scheme extends into integration. For instance, if $B$ invests in an ICO launched by $A$ and the token later loses value, $B$’s crypto is ``lost'' while $A$ obtains legitimate fiat proceeds, successfully off-ramping the crypto assets. Projects where developers abandon the protocol without making false promises are ``soft rug-pulls'' and may not be considered criminal~\cite{legge2024rugpulls}.
%\end{itemize}

\subsubsection{Consensus Protocols}
\label{subsubsec:consensus}

Recent studies reveal that consensus protocols---particularly Proof of Work (PoW) and Proof of Stake (PoS)---can be exploited for money laundering and sanctions evasion~\cite{team_cryptocurrency_2023,chainalysis2022,mandiantAPT43}. 
%\begin{itemize}

    \myparait{Placement via PoW.}
    Illicit fiat can be channeled into mining by investing on hardware, electricity, and hosting, generating fresh crypto that appears clean on-chain—a method reportedly used by the Iranian government to bypass sanctions~\cite{chainalysis2022crypto}. This approach converts fiat into blockchain-native assets, forming an unlinkable flow~(\S\ref{subsubsec:disconnect-fund-flow}). Another Chainalysis report identified mining pool addresses receiving both ransomware proceeds and legitimate earnings~\cite{team_cryptocurrency_2023}, evidencing ongoing \textit{commingling} activity (\S\ref{subsubsec:commingling}). 

    \myparait{Layering via PoW.}
    Mandiant discovered that North Korea’s \textit{Lazarus Group} launders stolen funds through cloud mining and hash rental services to acquire fresh, untainted assets~\cite{mandiantAPT43}, constituting a disconnection in fund flow~(\S\ref{subsubsec:disconnect-fund-flow}).

    \myparait{Layering via PoS.}
    Launderers may stake illicit funds to generate fresh rewards. The use of liquid staking derivatives (e.g., stETH) complicates tracing by embedding rewards into the asset’s value rather than paid as separate distributions~\cite{lido2021steth}. Staking via custodial or pooled DeFi services also facilitates commingling~(\S\ref{subsubsec:commingling}) of illicit funds~\cite{USTreasury2023DeFiRisk}. Launderers may also withdraw rewards to new wallets, re-stake them, or apply additional layers using other mentioned techniques.

%\end{itemize}

\subsubsection{Payment Channel Networks}
\label{subsubsec:intermediary-pcns}

Payment Channel Networks (PCNs) are off-chain protocols developed to improve blockchain scalability and efficiency~\cite{poon2016lightning,raiden_network,dziembowski2019perun}. They enable users to route payments through intermediaries without direct channels. For example, if Alice is connected to Bob and Bob to Charlie, Alice can pay Charlie via Bob using hashed timelock contracts (HTLCs)~\cite{kolachala2021sok}. Instead of recording each transaction on-chain, two parties open a bi-directional channel by locking funds in a multi-signature address, exchange rapid off-chain micro-transactions, and settle the net balance on-chain when the channel closes~\cite{kolachala2021sok,shojaeinasab_decoding_2024}.

\myparait{Layering via PCNs.}
Originally developed for scalability, PCNs have also emerged as effective obfuscation tools in the layering phase of money laundering. By fragmenting illicit funds and routing them through multiple off-chain paths—often via dormant or low-liquidity links—criminals can obscure transaction trails and complicate attribution~\cite{kolachala2021sok}. This tactic embodies both TFO and DFF (\S\ref{subsubsec:transaction-flow-obfuscation}, \S\ref{subsubsec:disconnect-fund-flow}). Tainted funds may be split across multiple paths and settled in different currencies, hindering flow reconstruction even when endpoints are observed~\cite{kolachala2021sok}. PCNs further enhance anonymity by recording only channel openings and closures on-chain; intermediate transfers remain off-ledger~\cite{shojaeinasab_decoding_2024}. While most PCNs are confined to a single blockchain—such as the \textit{Lightning Network} on Bitcoin~\cite{poon2016lightning} or the \textit{Raiden Network} on Ethereum~\cite{raiden_network}—their use of HTLCs and atomic swaps enables experimental cross-chain capabilities that facilitate chain-hopping~(\S\ref{subsubsec:chain-hopping}). Notably, frameworks like the \textit{Perun Network} propose support for cross-chain atomic swaps~\cite{perun_x_network}.

\section{Crypto Money Laundering Countermeasures}
\label{sec:ml-defenses}

In response to increasingly sophisticated crypto money laundering schemes, anti-money laundering (AML) measures have evolved along two paradigms: (a) detection measures, which aim to uncover laundering activity~(\S\ref{subsec:6.1-detective}); and (b) prevention measures, which seek to obstruct laundering before it occurs or prevent its recurrence~(\S\ref{subsec:6.2-preventive}).

\subsection{Detection Measures}
\label{subsec:6.1-detective}
Detection measures monitor and analyze transactional data to identify suspicious behavioral patterns across the blockchain ecosystem. Systematically, they fall into the following groups:

    \mypara{Rule-based detection.}
    AML detection has traditionally relied on rule-based systems, which use predefined thresholds and expert-crafted conditions to flag suspicious activity~\cite{vassallo2021application, pettersson2022combating}. Conditions such as transaction limits, unusual patterns, or specific event triggers can be used to generate alerts for investigation~\cite{vassallo2021application}. An advantage of rule-based frameworks is their interpretability and alignment with regulatory expectations, as institutions must justify flagged activity. Lin et al.~\cite{lin2023towards} used rule-based methods on Ethereum to trace laundering paths, while clustering techniques link addresses based on heuristics like multi-input usage~\cite{grinberg2012bitcoin, meiklejohn2013fistful}. Firms such as Elliptic~\cite{elliptic}, Chainalysis~\cite{chainalysis}, and CipherTrace~\cite{ciphertrace} apply these rules across chains for real-time AML risk assessment. However, rule-based systems are afflicted by high false positives, static rules that demand expert maintenance, and limited adaptability to evolving laundering tactics, making them less scalable without continuous expert updates~\cite{chen2018machine, vassallo2021application}.

    \mypara{Data-driven detection.}
    Machine learning techniques are used to inspect complex and evolving patterns in transactional data~\cite{vassallo2021application}. These automate detection, reducing manual rule reliance and enable adapting to evolving laundering tactics. Several approaches aim to classify licit and illicit transactions or entities~\cite{vassallo2021application}, often employing transaction graph analysis and addressing class imbalance challenges~\cite{vassallo2021application}. Other graph-based approaches~\cite{yu2023money,lin2024denseflow,song2024identifying,liu2023graph} model entities (e.g., users or wallets) as nodes, and relationships or flows as edges, based on the premise that laundering accounts form distinct structural patterns compared to legitimate ones. Another line of work~\cite{pham2016anomaly,monamo2016multifaceted} frames money laundering as an outlier detection problem, using clustering techniques to isolate anomalous users and transactions. Gu et al.~\cite{gu2022chain} focused on anomalous transaction amounts to flag potentially malicious exchanges. Other studies~\cite{toyoda2018multi,garin2023machine,zhou2022behavior} advanced illicit activity detection across services such as exchanges, mixers, darknet markets, gambling platforms, and mining entities by leveraging transaction histories, address data, and structural or behavioral features within Bitcoin and Ethereum. Further efforts analyzed account-level transaction histories to detect illicit addresses~\cite{farrugia2020detection,lin2019evaluation}, and combined transaction and account-level insights~\cite{elmougy2023demystifying} to enhance fraud detection.

\subsection{Prevention Measures}
\label{subsec:6.2-preventive}
Prevention efforts focus on regulatory controls, institutional safeguards, and public engagement, aiming to deter illicit activity by enhancing transparency. These include:

    \mypara{Regulatory deterrents.} 
    Global regulators, led by the FATF, have extended AML guidelines to \textit{Virtual Asset Service Providers} (VASPs)---including exchanges, stablecoin issuers, DeFi platforms, and NFT marketplaces~\cite{kolachala2021sok}. The U.S. and the European Union~(EU) have codified these into laws such as the \textit{Bank Secrecy Act} and the \textit{Fifth AML Directive}~(5AMLD), mandating KYC, \textit{Customer Due Diligence}~(CDD), transaction monitoring, and reporting of suspicious behaviors like smurfing~\cite{vassallo2021application}. KYC obliges VASPs to verify identities and flag suspicious activity. U.S. enforcement bodies such as FinCEN and the SEC implement these rules, even citing market manipulation risks to reject Bitcoin ETF applications~\cite{kolachala2021sok}. Consumer protection is also improving, with the EU’s MiCA and the UK’s FCA introducing safeguards, e.g., cooling-off periods for crypto purchases \cite{FCA2023coolingoff,mica}.
    
    \mypara{Promoting public awareness.} 
    Crypto money laundering operates as a complex socio-technical system that remains poorly understood by users~\cite{kasula2024awareness}. Launderers exploit this knowledge gap through fake ICOs~(\S\ref{subsubsec:icos}), in-person crypto trades~(\S\ref{subsubsec:disconnect-fund-flow}), and gift card sales~(\S\ref{subsubsec:atms-and-cards}), often targeting vulnerable groups such as students and the elderly~\cite{akartuna2022preventing,almeida2023review}. In response, governments, industry bodies, and academics have launched educational initiatives, including workshops and online awareness campaigns. Notable examples include programs led by the \textit{Central Bank of Bahrain}~(CBB), the \textit{Global Blockchain Business Council}~(GBBC)~\cite{yousif2024curbing}, and the \textit{Australian Securities and Investments Commission}~(ASIC)~\cite{ASIC2023}.
    
    \mypara{Blocklisting and sanctions.} 
    Blocklisting in cryptocurrencies involves identifying transactions linked to financial crimes and ``tainting'' the associated coins~\cite{kolachala2021sok}. This is feasible in cryptocurrencies like Bitcoin, where each UTXO’s publicly visible transaction history imparts a sense of uniqueness and non-fungibility. For example, the \textit{U.S. Office of Foreign Assets Control}~(OFAC) sanctioned Bitcoin addresses tied to Iranian ransomware operators in 2018~\cite{us2018Treasury}. Two primary blocklisting mechanisms have been proposed~\cite{moser2019effective,kolachala2021sok}: the \textit{Poison} policy, which taints all coins in subsequent transactions involving a blocklisted account, and the \textit{Haircut} policy, which limits tainting to the specific coins in the blocklisted transaction. Additionally, public ``taint-lists'' have been proposed to discourage engagement with tainted assets~\cite{yan2024shapley}.

    \mypara{Deanonymization and legal prosecution.}
    As noted in \S\ref{sec:ml-primer}, launderers exploit the inherently pseudonymous nature of the Web3 ecosystem, which collectively embolden their use of cryptocurrencies as a primary medium for illicit financial activity~\cite{alsalami2019sok,comolli_surfing_2021}. In response, government agencies and affiliated private organizations have invested in improving deanonymization techniques~\cite{deuber2022sok}. These efforts aim to unmask illicit actors, enabling subsequent legal investigation and prosecution. 
    However, PETs---mixers and AECs~(\S\ref{subsec:ml-intermediaries})---facilitate anonymity and financial secrecy, prompting technical~\cite{hong2018poster,wu2021towards} and regulatory responses. Centralized mixers like \textit{bestmixer.io} were dismantled by authorities, while decentralized ones, being non-custodial and jurisdiction-resistant, evade such actions~\cite{crawford2020knowing}. Despite this, OFAC sanctioned the decentralized mixer \textit{Tornado Cash} in 2022 for laundering \$7B~\cite{tornadoCash2022}. Privacy coins face similar scrutiny: Monero was delisted by Kraken across the \textit{European Economic Area}~(EEA) in 2024~\cite{kraken2024monero}, and will be further restricted under upcoming EU AML regulations---set to take effect by 2027--- banning anonymous accounts linked to privacy coins~\cite{ifcreview2025privacycoins}.

\section{Open Challenges \& Future Directions}
\label{sec:future-directions}
We outline core challenges in crypto AML alongside proposed research directions.

    \mypara{Data accessibility.}
    A key limitation of data-driven AML is the lack of large, reliably labeled datasets~\cite{vassallo2021application}. Despite curation, most labels remain approximate, degrading supervised model performance due to imbalance and limited coverage~\cite{lorenz2020machine}. Absence of ground truth raises false positives in detection and blocklisting, risking collateral damage. Among the data types in~\S\ref{sec:ml-primer}, on-chain data---though public---is vast, noisy, and ever-growing. Off-chain data face compliance barriers, often lacking KYC records or imposing access restrictions~\cite{comolli_surfing_2021}. Cross-chain traces inherit similar gaps or are entirely missing (e.g., DRT~\S\ref{subsubsec:disconnect-fund-flow}). Constructing transaction graphs entails fusing blockchain ledgers, smart contract logs, and proprietary VASP databases, complicating data correlation.

    \myparait{Future directions.}
    (a) Develop semi-/self-supervised models to reduce label reliance; (b) promote benchmark datasets via joint labeling by regulators and VASPs; (c) design graph aggregation frameworks for unified, temporally-aligned, multi-source representations; (d) expand data labeling efforts, building on prior contributions such as~\cite{baek2019model,monamo2016multifaceted,lorenz2020machine}.
%%%%%%%%%%%%%%%%%%%%%%%%%

    \mypara{Pseudonymity and decentralization.}
    Lack of custodianship limits legal authority to freeze assets or block addresses, while pseudonymity impedes deanonymization and prosecution. PETs strengthen anonymity, and their decentralized, non-custodial design renders dismantling infeasible. For instance, in 2024, a U.S. federal appeals court overturned OFAC's sanctions on Tornado Cash, holding that its immutable, autonomous smart contracts are not sanctionable entities~\cite{mayerbrown2024tornado}.

    \myparait{Future directions.}
    (a) Promote blockchain infrastructures which are regulatory-compliant, such as \textit{Ripple}---partnering with banks and payment providers to embed AML measures and KYC protocols into Web3~\cite{rippleCompliance}---and decentralized identity frameworks such as \textit{Worldcoin}~\cite{worldcoin2023whitepaper}, \textit{Ethereum Name Service}~(ENS)~\cite{ethereum-name-service} and \textit{Polygon ID}~\cite{polygonID}, that support verifiable user and service authorization; (b) Develop legal/technical frameworks enabling accountable PET deployment and develop compliance-aware PETs with selective disclosure~\cite{wagner2024seldom}.

    \mypara{Regulatory arbitrage.}
    The borderless nature of the crypto ecosystem undermines AML enforcement. VASPs based in lax regulatory environments can serve a global user base~\cite{comolli_surfing_2021} and used to move assets across non-compliant jurisdictions.

    \myparait{Future directions.}
    (a) Investigate cross-border enforcement mechanisms to mitigate fragmentation and propose incentive-aligned global AML standards for VASPs; (b) Smart contracts could embed logic to restrict access from addresses linked to sanctioned or non-compliant jurisdictions. This can be facilitated in dApps using \textit{compliance oracles} that inject off-chain regulatory data (e.g., KYC status, sanctions lists, AML risk scores) into smart contracts (e.g., \textit{Silent Data}~\cite{silentdata2025aml}).

    \mypara{Legitimate privacy concerns.} Deanonymization efforts further risk infringing the privacy of non-malicious users. PETs also safeguard legitimate interests—financial autonomy in authoritarian regimes, commercial confidentiality, etc. Their dual-use nature complicates regulation, as blanket bans risk infringing rights and deterring innovation.

    \myparait{Future directions.}
    The development of privacy-preserving mechanisms such as ZKP-based compliance frameworks (e.g, Silent Data), and identity systems (e.g, Polygon ID) offers a promising approach. These solutions enable verifiable, privacy-preserving identities and regulatory attestations, ensuring compliance while preserving user privacy.
%\end{itemize}

\section{Conclusion}
\label{sec:conclusion}

This survey aimed to address the absence of a systematized taxonomy of crypto money laundering schemes, by organizing existing and emerging strategies at an abstract level and examining their implementations across various mechanisms and infrastructures across the ecosystem. We highlighted the role of data-driven forensics and reviewed technical and operational countermeasures amid persistent challenges. Despite advances, launderers continue to adapt, casting a pressing need for scalable forensic tools and responsive policy frameworks to safeguard the integrity of the decentralized ecosystem.

\bibliographystyle{IEEEtran}
\bibliography{references.bib}
\balance

\appendices
\section{Literature comparison}
\label{appendix:literature-comparison}

\begin{table*}[t!]
\centering
\scriptsize
\renewcommand{\arraystretch}{1.1}
\newcommand{\angledheader}[1]{\rotatebox[origin=c]{90}{#1}}

\caption{Coverage of different dimensions of cryptocurrency laundering schemes across existing surveys.}
\begin{tabular}{c|*{6}{c}|*{11}{c}}
\toprule
\textbf{Surveys / Laundering Schemes} & \multicolumn{6}{c|}{\textbf{General Strategies}}
& \multicolumn{11}{c}{\textbf{Intermediary Mechanisms}} \\
\cline{2-7}\cline{8-18}
& \angledheader{\textbf{Transaction Flow Obfuscation}}
& \angledheader{\textbf{Commingling}}
& \angledheader{\textbf{Disconnected Fund Flow}}
& \angledheader{\textbf{Chain-hopping}}
& \angledheader{\textbf{Exploit Subjective Valuation}}
& \angledheader{\textbf{Exploit Regulatory Gaps}}
& \angledheader{\textbf{Mixers}}
& \angledheader{\hspace{5pt}\textbf{Anonymity-enhanced Cryptocurrencies}\hspace{5pt}}
& \angledheader{\textbf{Cryptocurrency Exchanges}}
& \angledheader{\textbf{Cross-chain Bridges}}
& \angledheader{\textbf{DeFi Mechanisms}}
& \angledheader{\textbf{NFTs}}
& \angledheader{\textbf{Cryptocurrency ATMs/Cards}}
& \angledheader{\textbf{Online Gambling}}
& \angledheader{\textbf{Initial Coin Offerings}}
& \angledheader{\textbf{Consensus Protocols}}
& \angledheader{\textbf{\quad Payment Channel Networks \quad}} \\
\midrule
Kolachala et al.~\cite{kolachala2021sok}           & \redcross & \redcross & \redcross & \redcross & \redcross & \redcross & \greencheck & \redcross & \redcross & \redcross & \redcross & \redcross & \redcross & \redcross & \redcross & \redcross & \greencheck \\
Lin et al.~\cite{lin2023towards}             & \greencheck & \redcross & \redcross & \greencheck & \redcross & \redcross & \greencheck & \greencheck & \greencheck & \redcross & \greencheck & \redcross & \redcross & \redcross & \greencheck & \redcross & \redcross \\
Shojaeinasab~\cite{shojaeinasab_decoding_2024} & \greencheck & \redcross & \greencheck & \greencheck & \redcross & \redcross & \greencheck & \greencheck & \greencheck & \redcross & \redcross & \redcross & \redcross & \redcross & \redcross & \redcross & \greencheck \\
Sneck~\cite{sneck2024cryptocurrency}    & \redcross & \redcross & \redcross & \greencheck & \redcross & \greencheck & \greencheck & \greencheck & \greencheck & \greencheck & \redcross & \greencheck & \greencheck & \greencheck & \redcross & \redcross & \redcross \\
Almeida et al.~\cite{almeida2023review}          & \redcross & \redcross & \greencheck & \greencheck & \redcross & \greencheck & \greencheck & \redcross & \greencheck & \redcross & \redcross & \redcross & \greencheck & \greencheck & \redcross & \redcross & \redcross \\
Comolli and Korver~\cite{comolli_surfing_2021}       & \redcross & \greencheck & \greencheck & \greencheck & \redcross & \greencheck & \greencheck & \greencheck & \greencheck & \redcross & \redcross & \redcross & \greencheck & \greencheck & \greencheck & \redcross & \redcross \\
Möser et al.~\cite{moser_inquiry_2013}         & \greencheck & \redcross & \greencheck & \redcross & \redcross & \greencheck & \greencheck & \redcross & \redcross & \redcross & \redcross & \redcross & \redcross & \redcross & \redcross & \redcross & \redcross \\
% Pending
Mooij~\cite{mooij2024regulating}        & \redcross & \redcross & \redcross & \redcross & \greencheck & \greencheck & \redcross & \redcross & \greencheck & \redcross & \redcross & \greencheck & \redcross & \redcross & \redcross & \redcross & \redcross \\\hline
\textbf{Ours}  & \greencheck & \greencheck & \greencheck & \greencheck & \greencheck & \greencheck & \greencheck & \greencheck & \greencheck & \greencheck & \greencheck & \greencheck & \greencheck & \greencheck & \greencheck & \greencheck & \greencheck \\
\bottomrule
\end{tabular}
\label{tab:laundering_sources}
\end{table*}

This appendix compares our systematization of laundering schemes with prior scholarly efforts that aim to classify money laundering in the cryptocurrency ecosystem. \autoref{tab:laundering_sources} (see next page) presents a taxonomy-based coverage analysis, highlighting the overlap and divergence between our work and existing surveys.

Kolachala et al.~\cite{kolachala2021sok} label their work as a systematization of knowledge paper on money laundering in cryptocurrencies; however, their primary focus is on money laundering countermeasures, particularly about blocklisting and traceability. Although they provide a detailed technical analysis of laundering via PCNs and HTLCs, they do not aim to propose a structured taxonomy of laundering strategies or intermediary mechanisms.

Lin et al.~\cite{lin2023towards} introduce a modified lifecycle-based taxonomy of money laundering stages in cryptocurrency systems. While this offers a broad conceptual framing, it leaves out many laundering schemes.

Shojaeinasab~\cite{shojaeinasab_decoding_2024} concentrates on forensic detection using address graph analysis and machine learning. The study briefly acknowledges key laundering infrastructures—mixers, AECs, exchanges, and PCNs, but does not attempt a formal categorization of laundering strategies or mechanisms. Its primary contribution lies in detection rather than systematization.

Sneck~\cite{sneck2024cryptocurrency} presents a systematic literature review of cryptocurrency-enabled laundering and terrorist financing. The work touches on several components of our taxonomy, including mixers, privacy coins, exchanges, chain-hopping, NFTs, and regulatory arbitrage. However, discussions included in the paper are example-driven and fail to formalize laundering mechanisms or attacker strategies. The absence of a structured taxonomy limits its utility for analytics or forensics.

Almeida et al.~\cite{almeida2023review} are among the few sources that attempt a structured classification of laundering methods by mapping them to PLI stages. They discuss several practical techniques which include in-person and P2P transactions, cryptocurrency ATMs, gambling, decentralized exchanges, and mixers. However, their categorization lacks granularity, omits many modern Web3 laundering mechanisms, and provides only brief descriptions of each tactic. Our work complements and extends their stage-based view by introducing a two-dimensional taxonomy encompassing both laundering intent and intermediary implementation.

Comolli and Korver~\cite{comolli_surfing_2021} provide a narrative overview of various laundering techniques and enforcement challenges in Web3. They refer to several elements also contained in our taxonomy such as commingling, mixers, jurisdictional arbitrage, and exchanges, while also touching on chain-hopping via atomic swaps and DEXs. However, their analysis does not differentiate strategies from mechanisms, nor does it cover emerging techniques such as cross-chain bridges, consensus participation, or NFT-based laundering. Our taxonomy refines and extends an unstructured list of laundering tools into a coherent framework.

Möser et al.~\cite{moser_inquiry_2013} explore transaction flow obfuscation patterns and the role of mixing services. Their study contributes early insights into disconnecting fund flow using swapping but does not aim to catalog laundering strategies or infrastructure comprehensively.

Mooij~\cite{mooij2024regulating} focuses specifically on money laundering risks in the metaverse, particularly involving NFTs. While the analysis is valuable for platform-specific vulnerabilities, it does not examine broader laundering schemes or mechanisms and is best understood as a complementary resource.

In contrast to all prior work, our study introduces a unified two-dimensional taxonomy that separates laundering strategies from their enabling mechanisms and maps them to the stages of the laundering lifecycle. We further identify the nature of forensic traces they leave---whether on-chain, off-chain, or cross-chain---enabling a more operationally useful classification. Our approach is grounded in qualitative thematic analysis and conceptual clustering, drawing from academic literature, regulatory reports, and forensic case studies. This allows us to capture under-explored techniques such as laundering via participation in mining and validating transactions, which are absent from mentioned sources.

\clearpage

\end{document}